\documentclass[reprint,nofootinbib,amsmath,amssymb,fontenc aps]{revtex4-1}
\usepackage{graphicx,dcolumn,bm,hyperref,float}
\usepackage{anyfontsize}

\newcommand{\be}{\begin{equation}}
\newcommand{\ee}{\end{equation}}
\newcommand{\beq}{\begin{equation}}
\newcommand{\eeq}{\end{equation}}
\newcommand{\ba}{\begin{eqnarray}}
\newcommand{\ea}{\end{eqnarray}}
\newcommand{\bef}{\begin{figure}}
\newcommand{\eef}{\end{figure}}

\newcommand{\Pol}{p}
\newcommand{\sq}{\beta}
\newcommand{\sqt}{S}
\newcommand{\sqs}{\zeta}
\newcommand{\kk}{ {\bf k} }
\newcommand{\sd}{\kappa}

\begin{document}


\title{Bound on Quantum Fluctuations in Gravitational Waves from LIGO-Virgo}

\author{$^1$Mark P.~Hertzberg}
\email{mark.hertzberg@tufts.edu}
\author{$^{1,2}$Jacob A.~Litterer}
\email{jacob.litterer@tufts.edu}
\affiliation{$^1$Institute of Cosmology, Department of Physics and Astronomy, Tufts University, Medford, MA 02155, USA
\looseness=-1}
\affiliation{$^2$Univ Coimbra, Faculdade de Ci\^encias e Tecnologia
	da Universidade \\ de Coimbra and CFisUC, Rua Larga, 3004-516 Coimbra, Portugal
\looseness=-1}

\begin{abstract}
We derive some of the central equations governing quantum fluctuations in gravitational waves, making use of general relativity as a sensible effective quantum theory at large distances. We begin with a review of classical gravitational waves in general relativity, including the energy in each mode. We then form the quantum ground state and coherent state, before then obtaining an explicit class of squeezed states. Since existing gravitational wave detections arise from merging black holes, and since the quantum nature of black holes remains puzzling, one can be open-minded to the possibility that the wave is in an interesting quantum mechanical state, such as a highly squeezed state. We compute the time and space two-point correlation functions for the quantized metric perturbations. We then constrain its amplitude with LIGO-Virgo observations. Using existing LIGO-Virgo data, we place a bound on the (exponential) squeezing parameter of the quantum gravitational wave state of $\zeta<41$.
\end{abstract}




\maketitle

\tableofcontents

\section{Introduction}

All the interactions of the Standard Model are known to arise from the behavior of quantum particles: photons, gluons, W/Z bosons, and Higgs bosons (e.g., see \cite{Weinberg}). There is clear experimental evidence for all these particles, along with all the fermions of the Standard Model. In contrast we do not have direct observations of the behavior of gravitons that underpin gravitation. Consistency between quantum mechanics and relativity implies that gravitons exist, and explains the structure of general relativity at long distances \cite{Weinberg:1964ew,Weinberg:1965rz,Deser:1969wk,Feynman,Hertzberg:2016djj,Hertzberg:2017abn,Hertzberg:2017nzl,Hertzberg:2020yzl,Hertzberg:2020gxu}. So it is a very worthwhile goal to search for observational consequences of quantum effects in gravitation. The inclusion of quantum effects can be done reliably at large distances, since general relativity is a well behaved effective theory for scales much larger than the Planck length. 

Recent observations of gravitational waves (GWs) by LIGO-Virgo are known to be broadly consistent with the predictions of classical general relativity \cite{LIGOScientific:2016aoc,LIGOScientific:2016lio,LIGOScientific:2016gtq,LIGO:2021ppb,McCuller:2021mbn}. Quantum corrections to the waves are ordinarily thought to be exceedingly small \cite{Rothman:2006fp,Dyson:2013hbl}. This is because the occupancy number of gravitons in a detectable wave is huge, and quantum corrections, or ``graviton shot noise", is suppressed. This conclusion is reliable under the assumption that the GW is in a coherent state, or similar, which are the most classical states. In such a case, we have essentially no chance to see any quantum behavior in the foreseeable future. 

On the other hand, GWs that are detectable have arisen due to black hole mergers (and neutron stars). The quantum character of black holes remains mysterious. So, while it is very plausible that the resulting GW is indeed in a coherent state, or similar, we can have an open mind to the possibility that the wave produced is in a much more striking quantum state, such as a highly squeezed state. In fact, squeezed states are naturally produced when quantum degrees of freedom are affected by a time dependent background (e.g. the production of squeezed state gravitons in the early universe, see Ref.~\cite{Grishchuk:1993ds}). This is indeed the case for the background when one considers quantum fluctuations in the gravitational field that are being acted on by the extreme spacetime evolution of merging black holes or neutron stars. Furthermore, it has been suggested that black holes may be highly quantum mechanical objects. 
Therefore while there is no known complete calculation of the production of squeezed state gravitons from mergers, as far as we are aware, it is worthwhile to first suppose squeezed states are produced and constrain the amount of squeezing that could have taken place in this process using existing data (we comment on important future work in Section \ref{Conc}). As the wave propagates from the merger to the earth, it is possible that a highly quantum state will not undergo appreciable decoherence, since gravitation is so weak. (There are related issues for dark matter, which may only have gravitational interactions \cite{Allali:2020ttz,Allali:2020shm,Allali:2021puy}).

In this work, we shall build the relevant formalism to describe GWs in squeezed states, compute their correlation properties, and perform a direct comparison to LIGO-Virgo data. We shall find a bound on the squeezing that seems rather weak; it only constrains a very large amount of squeezing. Nevertheless, if the merger produces squeezing at a rate that is analogous to how cosmic inflation produces extremely squeezed states \cite{Albrecht:1992kf,Polarski:1995jg}, then this may in fact be a useful bound. We note that in the context of cosmic inflation, this extreme ``squeezing" means the states are typically described as ``classical", since the field and momentum are closely correlated and matched well by stochastic methods. 
However, in the present context, this squeezing would render the observed gravitational waves from a merger {\em different} than the classical general relativistic prediction, so we refer to it as ``quantum" (irrespective of whether some stochastic method might be able to reproduce it).

There has been very important earlier work on the topic of quantum fluctuations in gravitation. In particular, a sequence of very significant works appears in Refs.~\cite{Parikh:2020nrd,Parikh:2020kfh,Parikh:2020fhy}. Here the authors calculate in detail how quantum fluctuations in gravitational waves lead to corresponding fluctuations in the arms of detectors. They provide the equations for the geodesic motion of the detectors, which include a kind of stochastic term, as well as a kind of radiation reaction term. These authors in Refs.~\cite{Parikh:2020nrd,Parikh:2020kfh,Parikh:2020fhy}, along with others in Refs.~\cite{Kanno:2020usf,Kanno:2021gpt}, also make significant progress in computing correlation functions of the gravitational field and the response. 
In this work, we build off the above foundational literature and provide a natural follow up.
Other interesting prior work includes Refs.~\cite{,Haba:2020jqs,Zurek:2020ukz,Verlinde:2019xfb,Guerreiro:2019vbq,Kuo:1993if,Ford:1994cr,Ford:1996qc,Coradeschi:2021szx,Cho:2021gvg,Guerreiro:2021qgk,Li:2022mvy}.  
In particular, our focus here is to compute specific details of the structure of a class of squeezed modes, the corresponding spacetime correlation functions, and a comparison to LIGO-Virgo data. After developing the necessary formalism, we compute correlations of squeezed quantum gravitational waves, whose standard deviation is then compared to the residual of existing data. This allows us to put a bound on the size of quantum fluctuations in observed gravitational waves.

The outline of this paper is the following: In Section \ref{Weak} we recap the form of the weak field metric for gravitational waves and the corresponding Hamiltonian. 
In Section \ref{QGW} we construct a family of wave functions, including squeezed states.
In Section \ref{CST} we determine their correlations in space and time.
In Section \ref{DCS} we analyze the detector response for a coherent state.
In Section \ref{DSS} we analyze the detector response for a squeezed state and place an observational bound on the squeezing parameter.
In Section \ref{Conc} we discuss.
In the Appendix we provide the details of the conceptually simpler case of the single harmonic oscillator for pedagogical purposes.

\section{Weak Field Hamiltonian}\label{Weak}

Let us begin with a quick review of the basics of weak gravitational fields.

\subsection{Metric for Gravitational Waves}

We will write the metric perturbation $h_{\mu\nu}$  around flat spacetime $\eta_{\mu\nu}$ as
\be
g_{\mu\nu}({\bf x},t)=\eta_{\mu\nu}+ h_{\mu\nu}({\bf x},t)
\ee
We will use units $c=1$, and our signature is (+,-,-,-).
Although the gravitational field may be extreme near the merging black holes, once it is near the earth, we know that it has entered the weak field regime, with metric fluctuations $h_{\mu\nu}$ that are small, i.e., $|h_{\mu\nu}|\ll 1$.

For gravitational waves propagating through the vacuum, we can go to transverse-traceless gauge in which the metric fluctuations $h_{\mu\nu}$ are found to take the form $h_{0\mu}=0$ and the spatial components we denote
$h_{ij}$. In this gauge, we have $\partial_i h_{ij}=0$ (transverse) and $\delta^{ij}h_{ij}=0$ (traceless).   

We will label the two gravitational wave polarizations by $\Pol=(+,\times)$.
For example, for a wave traveling in the $z$-direction, the spatial metric is of the form
\be
h_{ij}=\left(\begin{array}{ccc}
h_+(z,t) & h_\times(z,t) & 0 \\
h_\times(z,t) & -h_+(z,t) & 0 \\
0 & 0 & 0
\end{array}\right)
\ee

\subsection{Energy in Gravitational Waves}

Consider a wave traveling through space. In this gauge we can define the local energy density $\rho_{\mbox{\tiny{GW}}}({\bf x},t)$.
We can write this as a sum over the 2 polarizations $\Pol$ as
\be
\rho_{\mbox{\tiny{GW}}}({\bf x},t) = \sum_{\Pol=+,\times}{1\over 32\pi G}\left((\dot{h}_\Pol)^2+(\nabla{h}_\Pol)^2\right)
\ee

The total energy is given by integrating the energy density over space as 
\be
E=\int \!d^3x\,\rho_{\mbox{\tiny{GW}}}({\bf x},t)
\ee
By lifting the fields to operators, the corresponding Schr\"odinger equation is
\be
i\hbar{\partial\over\partial t}\Psi=\hat{H}\,\Psi
\ee
where the Hamiltonian operator $\hat{H}$ is equal to the above energy $E$ under the replacement to conjugate variables $(\hat{h},\hat{\pi})$
\be
h_p({\bf x}, t)\to\hat{h}_p({\bf x}, t),\,\,\,\,\,\,\dot{h}_p({\bf x}, t)\to(16\pi G)\,\hat{\pi}_p({\bf x}, t)
\ee
which obey standard (equal time) commutation relations
\be
[\hat{h}_p({\bf x}, t),\hat{\pi}_{p'}({\bf y}, t)]=i\,\delta_{pp'}\,\delta^3({\bf x}-{\bf y})\,\hbar
\ee

The corresponding probability density for a gravitational field configuration is $|\Psi|^2$.

\subsection{Energy in Each Mode }

It is convenient to discuss the properties of the waves in ${\bf k}$-space. Let us define the Fourier transform as
\be
\tilde{h}_{{\bf k},\Pol}(t) = \int\!d^3x\,h_p({\bf x},t)\,e^{-i{\bf k}\cdot{\bf x}}
\ee
By writing the energy in terms of the Fourier transform of $h_\Pol$, we have 
\be
E = \sum_{\Pol=+,\times}\int\!{d^3k\over(2\pi)^3}\,{1\over 32\pi G}\left(|\dot{\tilde{h}}_{{\bf k},\Pol}|^2+k^2|\tilde{h}_{{\bf k},\Pol}|^2\right)
\ee

It will also be convenient to define the theory in a finite size box of volume $V$. In this case, the modes become discrete. We can replace the integral over ${\bf k}$ by a discrete sum as
\be
\int\!{d^3k\over(2\pi)^3} \to {1\over V}\sum_{\bf k}
\ee
Then the energy can be written as the sum over modes and polarizations as
\be
E = \sum_{\bf k}\sum_{\Pol=+,\times} E_{{\bf k},\Pol} \label{Eg}
\ee
where 
\be
E_{{\bf k},\Pol} = {1\over V32\pi G}\left(|\dot{\tilde{h}}_{{\bf k},\Pol}|^2+k^2|\tilde{h}_{{\bf k},\Pol}|^2\right)\label{EGWk}
\ee
is the energy in each mode.

\section{Quantum Gravitational Wave}\label{QGW}

For a gravitational wave, the mathematics per mode is similar to the simple harmonic oscillator described in the Appendix. The reason being that the energy per mode of a gravitational wave of Eq.~(\ref{EGWk}) is of the same structure of the energy of a simple harmonic oscillator of Eq.~(\ref{ESHO}). The only difference is that the variable now is complex $\tilde{h}$, and we need to combine all the modes. (One can make direct comparison to the Appendix to provide further clarity, with the replacement $m \omega_0/2 \longrightarrow (32 \pi V G)^{-1} k$ when comparing the two systems.)

\subsection{Coherent and Squeezed States}

Of particular interest to us will be squeezed states, which can have enhanced fluctuations; while coherent states are the most classical, having minimal fluctuations that are too small to detect. Previously, Refs.~\cite{Parikh:2020nrd,Parikh:2020kfh,Parikh:2020fhy,Kanno:2020usf,Kanno:2021gpt,Guerreiro:2021qgk} made progress on studying these states. Our work here is to examine a particular class of squeezed states in more detail and its associated correlations. 

To find the squeezed state wave function for a field, it is useful to compute in Fourier space with $\tilde{h}_{{\bf k},p}$. Since this is complex, it is useful to decompose in terms of its real $\tilde{h}_{1{\bf k},p}$ and imaginary $\tilde{h}_{2{\bf k},p}$ parts as
\be
\tilde{h}_{{\bf k},p} = {\tilde{h}_{1{\bf k},p} + i\, \tilde{h}_{2{\bf k},p}\over\sqrt{2}},\,\,\,\,\,\,
\pi_{{\bf k},p} = {\pi_{1{\bf k},p} - i\, \pi_{2{\bf k},p}\over\sqrt{2}}
\ee

A form of the squeezed state wave function for each component is
\ba
\psi_s(h,t) \propto \prod_{a=1,2} \prod_{{\bf k}}\prod_{\Pol=+,\times} &&\exp\!\Big{[} i\,\epsilon_{a\kk,p} + {i\over2\hbar}\pi_{ac,\kk,p}\tilde{h}_{a\kk,p} \nonumber\\
&&\!\!\!\!\!\!\!\!\!\!\!\!- {k \, \sqt_{a\kk,p}(t) \over 64\pi V G \hbar}(\tilde{h}_{a{\bf k},\Pol}-\tilde{h}_{ac,{\bf k},\Pol}(t))^2\Big{]}\,\,\,\,\,\,\,\,\,\,\,  \label{psis}
\ea
(cf. Eq.~(\ref{psisApp})).

In this expression $\tilde{h}_{ac,{\bf k},p}(t)$ is a solution of the {\em classical} equation of motion, i.e.,
\be
\ddot{\tilde{h}}_{ac,{\bf k},\Pol}=-k^2 \tilde{h}_{ac,{\bf k},\Pol}
\ee
Note that back in position space, this is the classical wave equation
\be
\ddot{h}_{c,\Pol}=\nabla^2 h_{c,\Pol}
\ee
Note that if we take $\sqt_{a{\bf k},\Pol}(t)=1$ for all ${\bf k}$ and both polarizations, then this becomes a coherent state. 

More generally, we can include a family of squeezing functions $\sqt_{a{\bf k},\Pol}$ for each mode ${\bf k}$ and polarization $\Pol$, which can in principle be different between the real and imaginary parts $a$ (though we shall soon specialize to the case in which they are the same). By solving the Schr\"odinger equation, we find its time dependence is
\be
\sqt_{a{\bf k},\Pol}(t) = \mbox{Tanh}\!\left(\mbox{Tanh}^{-1}(\sq_{a{\bf k},\Pol}) + i\,k\,t\right)
\ee
which are specified by the choice of initial squeezing parameters
\be
\sqt_{a{\bf k},\Pol}(0) = \sq_{a{\bf k},\Pol}
\ee
which in principle can be different for each wave-vector ${\bf k}$ and each polarization $\Pol=+,\times$. The wave function also includes the function
\be
\epsilon_{a\kk,p}(t) = -{k\over 4} \int_0^t \sqt_{a\kk,p}(\tau)d\tau - {1\over 4\hbar} \tilde{h}_{ac,\kk,p}(t) \pi_{ac,\kk,p}(t)
\ee
to consistently include a $\beta\neq1$ and evolve according to the Schr\"odinger equation. Note in Eq.~(\ref{psis}) that the $\epsilon_{a\kk,p}(t)$ term only changes the phase of the wavefunction, and therefore drops out of the distribution $p(h,t)\propto|\psi_s(h,t)|^2$.

Importantly, the conjugate momentum $\pi_{\kk,p}$ is given from the Hamiltonian Eq.~(\ref{Eg}) as 
\be
\pi_{\kk,p} = {\partial E \over \partial \dot{\tilde{h}}_{\kk,p}} = {1\over 16\pi V G} \dot{\tilde{h}}^\dagger_{\kk,p}
\ee
and both obey the reality condition
\ba
\tilde{h}_{\kk,p} = \tilde{h}_{-\kk,p}^\dagger ~,~~~ \pi_{\kk,p} = \pi_{-\kk,p}^\dagger.
\ea
While the real and imaginary components obey
\ba
\tilde{h}_{1,\kk,p}&=&\tilde{h}_{1,-\kk,p},\,\,\,\,\,\,\tilde{h}_{2,\kk,p}=-\tilde{h}_{2,-\kk,p},\\
\pi_{1,\kk,p}&=&\pi_{1,-\kk,p},\,\,\,\,\,\,\pi_{2,\kk,p}=-\pi_{2,-\kk,p}. 
\ea

\subsection{Equal Time Fluctuations}

If we take the continuum limit (as we may always choose to do since the effective theory is only valid on large distances), then the distribution of the wavefunction Eq.$\,$(\ref{psis}) gives us the probability distribution for the field as $p(h,t)\propto|\psi_s(h,t)|^2$. For simplicity, let us report on results for a pair of identical squeezed functions $\sqt_{\kk,p}\equiv \sqt_{1\kk,p}=\sqt_{2\kk,p}$. In this case it is simplest to express results directly in terms of the complex fields $\tilde{h}_{{\bf k},\Pol}$. From taking the modulus 
squared of the above wave function, we have
\ba
&&p(h,t) \propto \nonumber\\
&&\prod_{\Pol=+,\times} \exp\!\left[-\int\!{d^3k\over(2\pi)^3}{k\,\mathbb{R}[\sqt_{{\bf k},\Pol}(t)] \over16\pi G\hbar}
|\tilde{h}_{{\bf k},\Pol}-\tilde{h}_{c,{\bf k},\Pol}(t)|^2\right]\,\,\,\,\,\,\,\,
\ea
where $\mathbb{R}[\sqt]$ is the real part of $\sqt$. 

The mean and variance are given by
\be
\langle \tilde{h}_{{\bf k},\Pol}\rangle=\tilde{h}_{c,{\bf k},\Pol}(t),\,\,\,\,\,\,
\sigma_{{\bf k},\Pol}^2= {V 8\pi G\hbar\over k} f_{{\bf k},\Pol}(t)
\ee
where
\be
f_{{\bf k},\Pol}(t)\equiv \sq_{{\bf k},\Pol}^{-1}\cos^2(k t)+\sq_{{\bf k},\Pol}\,\sin^2(k t)
\ee

In fact it is useful to define the departure from the mean as
\be
\delta h_\Pol\equiv h_\Pol-\langle h_\Pol\rangle
\ee
We can then form the 2-point correlation function in ${\bf k}$-space as
\be
\langle \delta \tilde{h}_{{\bf k},\Pol} \, \delta \tilde{h}_{{\bf k}',\Pol'}^* \rangle = {V 8\pi G\hbar\over k}f_{{\bf k},\Pol}(t)\,\delta_{{\bf k},{\bf k}'}\delta_{\Pol,\Pol'} 
\ee
where $\delta_{{\bf k},{\bf k}'}$ is the Kronecker delta function ($=1$ if ${\bf k}={\bf k}'$ and $=0$ if ${\bf k}\neq {\bf k}'$) and similarly for $\delta_{\Pol,\Pol'}$.

In the continuum limit we can write this in terms of the Dirac delta function $\delta^3({\bf k}-{\bf k}')$ as
\be
\langle \delta \tilde{h}_{{\bf k},\Pol} \, \delta \tilde{h}_{{\bf k}',\Pol'}^* \rangle =(2\pi)^3\,\delta^3({\bf k}-{\bf k}')\,\delta_{\Pol,\Pol'}\,P_{\Pol}({\bf k},t)
\label{2ptpower}\ee
where the ``power spectrum" is given by
\be
P_{\Pol}({\bf k},t)=P_v(k)\,f_{{\bf k},\Pol}(t)
\ee
where
\be
P_v(k)={8\pi G\hbar\over k}
\label{Pvac}\ee
is the power spectrum of vacuum fluctuations.

It is important to return to position space, since we are ultimately interested in the motion of mirrors in interferometers which are well localized in position space.
By taking the inverse Fourier transform we obtain
\be
\langle \delta h_\Pol({\bf x},t)\,\delta h_{\Pol'}({\bf y},t)\rangle = \delta_{\Pol,\Pol'}\int\!{d^3 k\over(2\pi)^3}\,P_{\Pol}({\bf k},t)\,e^{i{\bf k}\cdot({\bf x}-{\bf y})}
\ee
Although it can be useful to see correlations in space at a fixed time, the interferometer experiments are sensitive to something else. In particular, one watches a mirror swing back and forth over time. Therefore it is important to understand correlations in time as well, as we now compute. 

\section{Correlations in Space and Time}\label{CST}

Here we report on the correlation functions at different times and space. It is convenient to perform the calculation in Fourier space, where we find an extension of Eq.~(\ref{2ptpower}) to
\be
\langle \delta \tilde{h}_{{\bf k},\Pol}(t) \, \delta \tilde{h}_{{\bf k}',\Pol'}^*(t') \rangle =(2\pi)^3\,\delta^3({\bf k}-{\bf k}')\,\delta_{\Pol,\Pol'}\,Q_{\Pol}({\bf k},t,t')
\ee
where the mixed time power spectrum $Q$ is given by
\be
Q_{\Pol}({\bf k},t,t')=P_v(k)\,F_{{\bf k},\Pol}(t,t')
\ee
where $P_v$ is the power spectrum of vacuum fluctuations Eq.~(\ref{Pvac}). By using the Heisenberg equation of motion (an illustrative example is given in the Appendix), we find that $F$ is given by
\ba
F_{{\bf k},\Pol}(t,t')&\equiv& \sq_{{\bf k},\Pol}^{-1}\cos(k t)\cos(k t')+\sq_{{\bf k},\Pol}\,\sin(k t)\sin(k t')\nonumber\\
&&+i\,\sin(k(t'-t))
\ea

Returning to position space, this becomes
\ba
\xi_p({\bf x},{\bf y},t,t') & = & \langle h_p({\bf x},t)\,h_p({\bf y},t')\rangle -\langle h_p({\bf x},t)\rangle\langle h_p({\bf y},t')\rangle \,\,\,\,\, \nonumber\\
& = & \delta_{\Pol,\Pol'}\int\!{d^3 k\over(2\pi)^3}\,Q_{\Pol}({\bf k},t,t')\,e^{i{\bf k}\cdot({\bf x}-{\bf y})}
\ea

\subsection{Monochromatic Squeezing Function}

If, for simplicity, we assume there is only a single mode ${\bf k}^*$ that is significantly squeezed, then we can write 
\be
\sq_{{\bf k},\Pol}=1+{e^{2\sqs_\Pol}k^{*3}\!\over2}\,(2\pi)^3(\delta^3({\bf k}-{\bf k}^*)+\delta^3({\bf k}+{\bf k}^*))
\label{betadelta}\ee
where $\sqs_\Pol$ is the {\em dimensionless} strength of the squeezing. Note that we added a pair of delta-functions to ensure the reality condition is obeyed, i.e., we need $\sq_{{\bf k},\Pol}=\sq_{-{\bf k},\Pol}^*$. Also note that we added 1 to every mode, which corresponds to no-squeezing. 

With this form for $\beta$, we have
\ba
&&\xi({\bf x},{\bf y},t,t') \nonumber\\
&=& \delta_{\Pol,\Pol'}{1\over2\pi^2}\int_0^\infty dk\,k^2\,P_v(k){\sin(k|{\bf x}-{\bf y}|)\over k|{\bf x}-{\bf y}|} e^{-ik(t-t')}\nonumber\\
&+&\delta_{\Pol,\Pol'}{e^{2\sqs_\Pol}k^{*3}\!\over2}\!\int\! d^3k\Big[P_v(k)(\delta^3({\bf k}-{\bf k}^*)+\delta^3({\bf k}+{\bf k}^*))\nonumber\\
&&\,\,\,\,\,\,\,\,\,\,\,\,\,\,\,\,\,\,\,\,\,\,\,\,\,\,\,\,\,\,\,\,\,\,\,\,\,\,\,\,\,\,\,\,
\times\sin(kt)\sin(kt')e^{i{\bf k}\cdot({\bf x}-{\bf y})}\Big]\,\,\,\,\,\,\,\,\,
\ea
We can then use the following fact
\be
\int_0^\infty dk\,k^2\,{1\over k}{\sin(k r)\over k r}e^{-ik(t-t')} = {1\over r^2-(t-t')^2}
\ee
and we can trivially carry out the delta-function integrals. 
Hence the full 2-point correlation function is
\ba
&&\xi_p({\bf x},{\bf y},t,t') = \delta_{\Pol,\Pol'}{1\over 2\pi^2}{8\pi G\hbar\over|{\bf x}-{\bf y}|^2-(t-t')^2}\nonumber\\
&&+\delta_{\Pol,\Pol'}8\pi G\hbar \,e^{2\sqs_\Pol}k^{*2}\sin(k^*t)\sin(k^*t')\cos({\bf k}^*\!\cdot\!({\bf x}-{\bf y}))\,\,\,\,\,\,\,\,\,\,
\label{2ptauto}\ea
Note the sinusoidal oscillations in both $t$ and $t'$; this is a property of the very restrictive monochromatic squeezing; this will be altered when we move to more realistic squeezing functions, as we turn to now.

\subsection{Smoothed Out Squeezing Function}

Let us consider squeezing a range of modes, rather than only a unique value ${\bf k}^*$. Since a gravitational wave involves a continuum of modes, this seems more realistic, and we may retain the feature of squeezing a unique primary mode to be later identified with an observed gravitational wave peak frequency. Suppose the wave is heading in the (positive) $z$-direction, with its mean wavenumber of ${\bf k}^*=k^*\hat{z}$ and standard deviation $\sd$. We smear out Eq.~(\ref{betadelta}) to become
\ba
\sq_{{\bf k},\Pol}=1\,+\,&&{e^{2\sqs_\Pol}k^{*2}k\!\over2}\,(2\pi)^3\delta(k_x)\delta(k_y)\times\nonumber\\
&&{1\over\sqrt{2\pi\sd^2}}\left[e^{-(k_z-k^*)^2/2\sd^2}+e^{-(k_z+k^*)^2/2\sd^2}\right]\,\,\,\,\,\,\,\,\,
\label{betasmooth}\ea
Note that we also smoothed out the prefactor $k^{*3}\to k^{*2}k$ for convenience (for narrow smoothing, the correction is small, but the integrals become simpler with this choice).  By carrying out the above integrals, and taking the large $t$ and $t'$ limit, we obtain
\ba
&&\xi_p({\bf x},{\bf x}',t,t') = \delta_{\Pol,\Pol'}{1\over 2\pi^2}{8\pi G\hbar\over|{\bf x}-{\bf x}'|^2-(t-t')^2}\nonumber\\
&&+\delta_{\Pol,\Pol'}2\pi G\hbar \,e^{2\sqs_\Pol}k^{*2}\times\nonumber\\
&&\sum_{\mp}e^{-((z-z')\mp(t-t'))\sd^2/2}\cos(k^*((z-z')\mp(t-t')))\,\,\,\,\,\,\,\,\,
\label{2ptautoSmooth}\ea
Note that having performed the smoothing, we have removed the oscillations in Eq.~(\ref{2ptauto}) and obtained a result with time translation invariance, being only a function of $t-t'$. Since the physical scenario we have in mind is an earth based detector, with gravitational waves sourced by a merger hundreds of millions of lightyears away, we may always take the late time limit. In this limit, the above result is intuitively more physically reasonable than Eq.~(\ref{2ptauto}) with monochromatic squeezing. A version whose fluctuations ``hit" in the region of the classical wave itself can be obtained, as we now discuss.

\subsection{Classical Wave Modulation}

A more realistic smoothing function should lead to the second line in Eq.~(\ref{2ptautoSmooth}) being appreciable at the center of the classical wave packet $h_c$, whose state one is taking to be squeezed. This can be accommodated by taking this modulation to be adiabatic, i.e., to suppose that the above high frequency modes are modulated by a relatively low frequency mode, whose corresponding wavelength is of the order of the size of the classical wave packet. So in this adiabatic approximation the second line should be re-scaled by an overall factor $\mu_\Pol(z,t)$, with
\beq
\mu_\Pol(z,t) = \Bigg\{\begin{array}{c}
1,\,\,\,\, |z-t-\phi_c|\ll\lambda_c\\
0,\,\,\,\,|z-t-\phi_c|\gg\lambda_c\end{array}
\eeq
where $\phi_c$ is the phase of the center of the classical wave packet (i.e., $h_c(z=t+\phi,t)$ is large) and $\lambda_c$ is the overall size of the wavepacket. An example modulation function could be $\mu_\Pol(z,t)=\exp(-(z-t-\phi_c)^2/\lambda_c^2)$.

\section{Detector Response for Coherent State}\label{DCS}

Let us first focus on the most classical possible state, i.e., a coherent state. So in this section we set
\be
\sq_{{\bf k},\Pol}=1
\ee
(we consider $\sq_{{\bf k},\Pol}\neq1$ in the next section).

Now despite the fact that this is essentially the ``most classical" state, nevertheless, the above result seems to indicate that as ${\bf x}\to{\bf y}$ the quantum fluctuations become infinite!  However, we should note that this is only true if we really integrate the wave-numbers all the way up to $k\to\infty$. But this is unrealistic. The reason is that any detector, such as LIGO-Virgo, cannot resolve arbitrarily high frequencies. The frequencies of the wave are related to the wave-number by
\be
f={\omega\over 2\pi}={k\over 2\pi}
\ee
Let's introduce a ``response" function $R(k)$, defined such that $R=1$ when the detector can resolve easily and $R=0$ when the detector cannot. We can insert this into our above result as follows
\be
\xi({\bf x},{\bf y},t,t')_R = \delta_{\Pol,\Pol'}\int\!{d^3 k\over(2\pi)^3}\,P_v(k)\,e^{i{\bf k}\cdot({\bf x}-{\bf y})-ik(t-t')} R(k)
\ee
where the subscript ``R" notation indicates we take into account the detector response.

\begin{figure}[t!]
\centering
\includegraphics[width=\columnwidth]{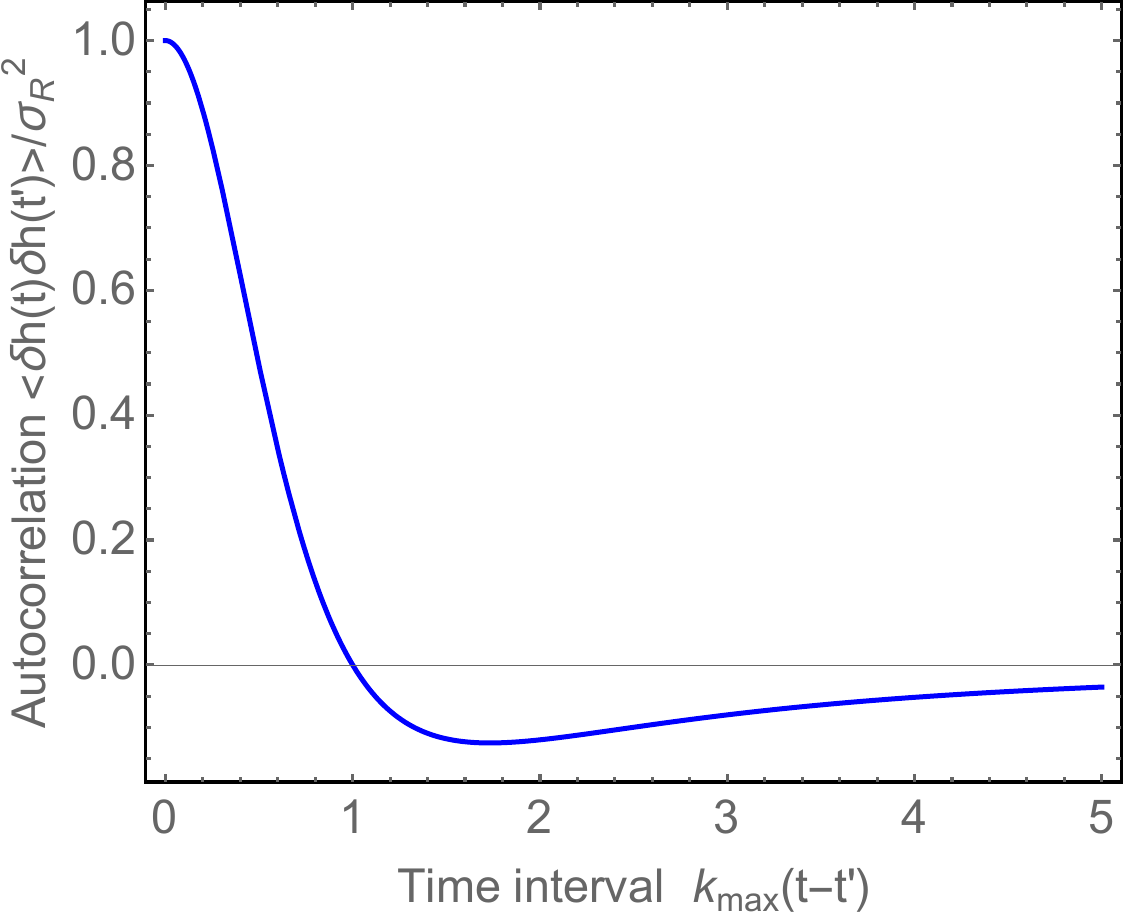}
\caption{(Normalized) autocorrelation function of vacuum fluctuations, defined with some cut off $k_{max}$.}
\label{AutoCorrelation}
\end{figure}

If we consider the variance of fluctuations at a single spacetime point ${\bf x}\to {\bf y}$ we obtain the ``autocorrelation" function
\be
\langle \delta h_\Pol(t)\delta h_\Pol(t')\rangle_R = \int\!{d^3 k\over(2\pi)^3}\,P_v(k)e^{-ik(t-t')}\,R(k)
\ee
A simple choice to suppress high frequency modes is
\be
R(k)
=\exp\left[-{k\over k_{max}}\right]
\ee
where $k_{max}$ sets the maximum characteristic wavenumber the detector can resolve. Carrying out the above integral with this $R(k)$ leads to
\be
\langle \delta h_\Pol(t)\delta h(t')\rangle_R =
{4\over\pi}\left(k_{max}\over \omega_{Pl}\right)^2{1\over(1+i\,(t-t')\,k_{max})^2}
\ee
where we have introduced the ``Planck frequency"
\be
\omega_{Pl}\equiv{1\over\sqrt{G\hbar}}\approx 1.9\times10^{43}\,\mbox{sec}^{-1}
\ee
Note that in general this autocorrelation function is complex, as the above is not a hermitian operator. However, to connect to a physical observable, we can symmetrize over the arguments to define a hermitian operator as follows
\beq
\xi({\bf x},{\bf y},t,t')_S\equiv (\xi({\bf x},{\bf y},t,t')_R+\xi({\bf y},{\bf x},t',t)_R)/2
\eeq
This gives
\be
\langle \delta h_\Pol(t)\delta h(t')\rangle_S = {4\over\pi}\left(k_{max}\over \omega_{Pl}\right)^2
{1-k_{max}^2(t-t')^2\over(1+ k_{max}^2(t-t')^2)^2}
\ee
The standard deviation $(t\to t'$) in the fluctuations is therefore
\be
\sigma_R=\sqrt{\langle (\delta h_\Pol)^2\rangle_R}=\sqrt{4\over\pi}\left(k_{max}\over \omega_{Pl}\right)
\ee
A plot of this symmetrized autocorrelation function is given in Figure \ref{AutoCorrelation}.

\subsection{Detector Limits and Comparison}

In the LIGO-Virgo detector, the maximum frequency that the interferometer can respond to reliably is on the order of
\be
{k_{max}\over2\pi}=f_{max}=\mathcal{O}(10^3)\,\mbox{Hz}
\ee
This gives a standard deviation in quantum fluctuations on the order of 
\be
\sigma_R =\mathcal{O}(10^{-40})
\ee
We should compare the size of these quantum fluctuations to the size of the classical gravitational waves detected at LIGO-Virgo from merging binary black holes. From the LIGO-Virgo paper, 
we see that the measured amplitude of the wave is 
\be
h_c =\mathcal{O}(10^{-21})
\ee
Putting this altogether we see that the relative size of the quantum fluctuations in a coherent state to the classical value is
\be
{\sigma_R\over h_c}=\mathcal{O}(10^{-19})
\ee
This is far too small to detect; as previously noted \cite{Parikh:2020nrd,Parikh:2020kfh,Parikh:2020fhy}. The current capability of LIGO-Virgo to detect fluctuations is about an order of magnitude below the classical value from mergers; certainly it cannot detect 19 orders of magnitude below. 

For the {\em pure Minkowski vacuum state} fluctuations, there is a theoretical question of whether such fluctuations are detectable even in principle. More appropriately, since the world is actually built out of particles, one needs to construct the {\em dressed state} of the detector. Once this is constructed, it is unclear that any directly physical consequences of the vacuum fluctuations remain. After all, one needs to study interactions {\em between} materials (like the interactions between plates in the Casimir effect) to see physical consequences. This point seems to have been missed in some previous analyses of gravitational vacuum fluctuations in Minkowski space in the literature. 

On the other hand, for the {\em coherent state}, this should manifest itself as physical shot noise. In any case, since the effect is so small, we do not pursue this further here.

\section{Detector Response for Squeezed State}\label{DSS}

Since the coherent state fluctuations are far too small, we can turn to the squeezed state, with strength of squeezing parameter $\sqs_\Pol$. 

In this case we only need to pay attention to the second term in Eq.~(\ref{2ptauto}) or Eq.~(\ref{2ptautoSmooth}). 
Then the fluctuations are well behaved as ${\bf x}\to{\bf x}'$ and we expect to resolve these modes, which only apply for ${\bf k}={\bf k}^*$ since that is already anticipated to be a mode of the classical wave. 

\begin{figure}[t!]
\centering
\includegraphics[width=\columnwidth]{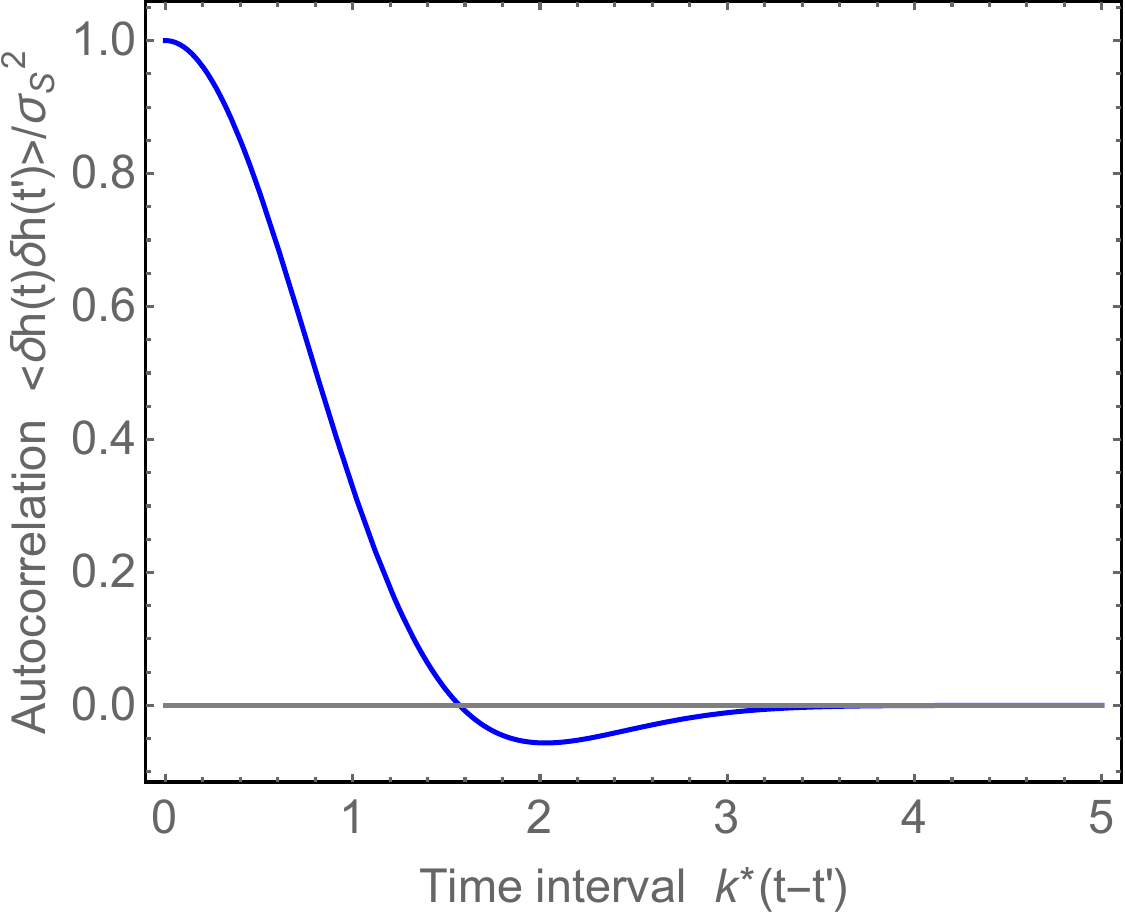}
\caption{(Normalized) autocorrelation function for squeezed modes from a smoothed distribution. Here we took $\sd=k^*$ and evaluated $z$ at $t+\phi_c$; the center of the wave packet.}
\label{AutoCorrelation2}
\end{figure}

Focussing our attention on the case of the smooth squeezing function, the fluctuations are then
\be
\langle \delta h_\Pol(t) \delta h_\Pol(t')\rangle
=A\,e^{-(t-t')\sd^2/2}\cos(k^*(t-t'))\,\mu_\Pol(z,t)\,\,\,\,\,\,\,\,\,
\label{2ptimeautoSmooth}\ee
where the overall amplitude is defined as
\be
A=4\pi G\hbar\, e^{2\sqs_\Pol}k^{*2}
\ee
At the center of the classical wave packet, where $\mu_\Pol\to1$, the remaining shape 
exhibits the nice features of time translation invariance, as it is only a function of $t-t'$.
The corresponding correlation is shown in Figure \ref{AutoCorrelation2}. 

\subsection{Constraints from LIGO-Virgo}

Let us consider the amplitude of fluctuations at each moment in time. To do so, we take $t\to t'$, and we evaluate $\mu_\Pol\to1$ as we are interested in the maximum fluctuation. We wish to compare this to the noise seen at LIGO-Virgo.

Writing the wave-number in terms of the frequency of the wave $k^*=2\pi f^*$, sending $t\to t'$, and taking a square root gives the standard deviation $\sigma_S =\sqrt{\langle (\delta h_\Pol)^2\rangle}$ in the squeezed state of
\be
\sigma_S = \sqrt{4\pi} \,e^{\sqs_\Pol}\!\left(2\pi f^*\over \omega_{Pl}\right)
\label{sigmaS}\ee
The scaling here is broadly consistent with the scalings estimated in prior works \cite{Parikh:2020kfh,Parikh:2020fhy,Kanno:2020usf,Guerreiro:2021qgk}.

Let us compare this to the first observation of merging black holes seen by LIGO-Virgo GW150914 \cite{LIGOScientific:2016aoc}. As is well known, the response by the interferometer is in good agreement with the predictions of classical general relativity. There does exist a residual noise, i.e., a residual difference between observation and theoretical prediction. Using data from the event GW150914, we computed the standard deviation of the residual from LIGO-Virgo at Hanford (H) and Livingston (L), which are roughly equal
\be
\sigma_{H}\approx\sigma_L\approx 0.16\times 10^{-21} .
\ee
Very reasonably, this residual can be mostly accounted for from various well known effects, such as thermal noise, photon shot noise, etc. So at the very least we can use this as a definite {\em upper bound} on the size of the quantum gravitational fluctuations $\sigma_S$ in our squeezed state. To evaluate $\sigma_S$ as above, we need a value of $f^*$, a central frequency of the (classical) wave. From GW150914 data, the frequencies of largest support in the wave occurred for
\be
f^*\sim 200\,\mbox{Hz}
\ee
By inserting this value into Eq.~(\ref{sigmaS}) and demanding 
$\sigma_S<\sigma_H\approx\sigma_L$,  
we obtain a bound on $\zeta_\Pol$ of
\be
\zeta_\Pol<41
\ee
(for each mode) which is our primary finding. By estimating residuals from known effects, one could improve this bound a little. Furthermore, one could analyze more carefully the ringdown and/or inspiral phases. We leave these considerations for future work.

\subsection{Temporal Correlations}

The existing LIGO-Virgo data does not exhibit any known unexplained correlation. So drawing from these correlated fluctuations, the bound should approximately reproduce the $\zeta_p\lesssim 41$ found above. The reason being that the above correlations in the squeezed state fall off beyond the inverse characteristic frequency of the signal (see Figure \ref{AutoCorrelation2}), so these are rather short ranged correlations. Nevertheless for further precision, one can run the autocorrelation function on the data to potentially improve the bound on $\zeta_p$ marginally.

\section{Discussion}\label{Conc}

At large distances, gravity is a consistent quantum effective theory. In this work we have computed the properties of a gravitational wave in a quantum mechanical squeezed state. We paid particular attention to the two-point correlation function of the gravitational wave in both space and time. By considering a smooth range of modes, we obtained the correlation function in Eq.~(\ref{2ptautoSmooth}) (along with the modulation function $\mu_\Pol$), in which the fluctuations are nicely separated into a vacuum piece, which is not directly measurable, and an enhanced squeezed piece, which could in principle correct the motion of detectors in an interferometer. By using existing LIGO-Virgo data of event GW150914, we placed a bound on the (exponential) squeezing parameter of $\zeta<41$. Although this is a relatively weak bound, it is interesting to be able to use existing data to place real constraints. Further detailed constraints from combining all LIGO-Virgo data of its many events is useful future work.

An important topic for future work is to compute from first principles the actual state set up by the merger of black holes. The usual assumption is that the state is close to classical, i.e., close to a coherent state, in which case the quantum fluctuations are predicted to be negligibly small. In this case, a future detection of quantum fluctuations would be unexpected and of profound significance as it would render gravitation incompatible with standard quantum mechanics with a universal uncertainty bound. Alternatively, if the black hole merger creates huge squeezing and/or if the individual black holes are intrinsically very quantum, as some speculative models have suggested, the state could be highly squeezed. It would then also become important to know what configurations of black holes and mergers produce the most squeezing. This all deserves further analysis. To complement this theoretical work, future work should also improve upon the statistical method of this paper in order to produce more rigorous observational bounds using all of the available gravitational wave data.
We are now in the era of gravitational wave astronomy, with several detectors, including LIGO, Virgo, GEO600, KAGRA.  So any information that we can use to glean even the smallest clues about quantum gravity may be useful. 

\section*{Acknowledgments}
M.~P.~H is supported in part by National Science Foundation grant PHY-2013953.
J.~A.~L is supported by a postdoctoral research fellowship under the FCT, I.P. grant No. CERN/FIS-PAR/0027/2021 and partially by the CFisUC project No. UID/FIS/04564/2020.
We would like to thank Eunice Beato, Andi Gray, and Erin Wilson for discussion.

\section*{Appendix: Simple Harmonic Oscillator}\label{AppSHO}

This appendix is for pedagogical purposes; for any reader who is new to this subject. By understanding the results here for the simple harmonic oscillator, one can extend them to the case of a quantum gravitational wave, as we did in the main part of the paper.

The standard simple harmonic oscillator is a body of mass $m$, oscillating on a spring with natural frequency $\omega_0$. The energy is a sum of kinetic and potential energy as
\be
E_{\mbox{\tiny{SHO}}} = {1\over2}m\dot{x}^2+{1\over2}m\omega_0^2 x^2\label{ESHO}
\ee
But this basic structure appears much more generically, including the form seen earlier for each mode of a gravitational wave in Eq.~(\ref{EGWk}).

\subsection{Ground State}
Classically, a harmonic oscillator would sit at rest at the bottom of the potential with minimal energy of zero. But this is not allowed by the Heisenberg uncertainty principle. Instead the quantum ground state wave function is given by 
\be
\Psi_g(x,t)\propto \exp\!\left[-{i}E_0 t/\hbar-{1\over2\hbar}m\omega_0 x^2\right]
\ee
where $E_0={1\over2}\hbar\omega_0$ is the ground state energy.

The probability density distribution for where the particle can be found is
\be
\rho_g(x,t)=|\Psi_{g}(x,t)|^2\propto \exp\!\left[-{1\over\hbar}m\omega_0 x^2\right]
\ee
This is a type of Gaussian distribution as it fits the standard form
\be
\rho(x,t)\propto \exp\!\left[-(x-\langle x\rangle)^2/(2\sigma^2)\right]
\ee
Here the mean and variance of position is
\be
\langle x\rangle = 0,\,\,\,\,\,\,\sigma_x^2={\hbar\over 2m\omega_0}
\ee
While the mean and variance of momentum is
\be
\langle p\rangle = 0,\,\,\,\,\,\,\sigma_p^2={\hbar m\omega_0\over 2}
\ee
Note that the product of uncertainties is
\be
\sigma_x\,\sigma_p={\hbar\over2}
\ee
which is the minimum value allowed by the Heisenberg uncertainty principle.

\subsection{Coherent State}
A coherent state is considered the ``most classical state": Like the ground state, it minimizes the Heisenberg uncertainty principle at all times. However, unlike the ground state, it also has a mean value that undergoes familiar classical motion.

The wave function is given by
\be
\psi_c(x,t)\propto  \exp\!\left[i\,\epsilon(t)+ip_0(t)x/\hbar-{1\over2\hbar}m\omega_0 (x-x_0(t))^2\right]
\ee
where the phase is given by
\be
\epsilon(t)=-{1\over2}\omega_0\,t-{1\over2\hbar}x_0(t)p_0(t)
\ee
Here $x_0(t)$ is a solution of the {\em classical} equation of motion, i.e.,
\be
\ddot{x}_c=-\omega_0^2 x_c
\ee
Its solutions are given by
\be
x_0(t) = A \cos(\omega_0t-\varphi) \label{x0}
\ee
where $A$ is the amplitude of oscillation and $\varphi$ is the phase. Also, $p_0(t)=m\dot{x}_0(t)$ is the classical momentum.

The corresponding probability distribution that is in fact the same as the ground state, except it is displaced as follows
\be
\rho_c(x,t)\propto \exp\!\left[-{1\over\hbar}m\omega_0 (x-x_0(t))^2\right]
\ee

Hence the coherent state is also a Gaussian distribution with mean and variance
\ba
&&\langle x\rangle = x_0(t),\,\,\,\,\,\,\sigma_x^2={\hbar\over 2m\omega_0}\\
&&\langle p\rangle = p_0(t),\,\,\,\,\,\,\sigma_p^2={\hbar m\omega_0\over 2}
\ea
So just like the ground state, it too saturates the uncertainty principle
\be
\sigma_x\,\sigma_p={\hbar\over2}
\ee
while allowing for large oscillations. This makes it especially classical. 

\subsection{Squeezed State}
A squeezed state is similar to a coherent state, except that the product of variances does {\em not} saturate the uncertainty principle limit, nor is it time independent. The initial wave function is taken to be 
\be
\psi_s(x,0)\propto  \exp\!\left[{i\over\hbar}p_0(0)x-{\sq\over2\hbar}m\omega_0 (x-x_0(0))^2\right]
\ee
where $\sq$ is the ``squeezing parameter". If $\sq=1$ then this returns to the above coherent state. While for $\sq\neq1$ we have a so-called squeezed state. By solving the time dependent Sch\"odinger equation, the time evolved state can be shown to be
\be
\psi_s(x,t)\propto  \exp\! \Big{[}
i\,\epsilon(t)+{i\over\hbar}p_0(t)x-{\sqt(t)\over2\hbar}m\omega_0 (x-x_0(t))^2\Big{]} \label{psisApp}
\ee
Here the function $\epsilon(t)$ can be represented as
\be
\epsilon(t)=-{1\over2}\omega_0\!\int^t_0 dt'\sqt(t')-{1\over2\hbar}x_0(t)p_0(t)
\ee
Importantly, we now have the time evolved squeezing function $S(t)$. In general it is complex valued; it found to be given by
\be
\sqt(t) = \mbox{Tanh}\!\left(\mbox{Tanh}^{-1}(\sq) + i\,\omega_0\,t\right) \label{S}
\ee
Note that initially ($t=0$) we have
\be 
\sqt(0)=\sq
\ee

The probability distribution is
\be
\rho_s(x,t)\propto \exp\!\left[-{\mathbb{R}[\sqt(t)]\over\hbar} m\omega_0(x-x_0(t))^2\right]
\ee
where $\mathbb{R}[\sqt(t)]$ means the real part of $\sqt(t)$.

\begin{figure}[b!]
\centering
\includegraphics[width=\columnwidth]{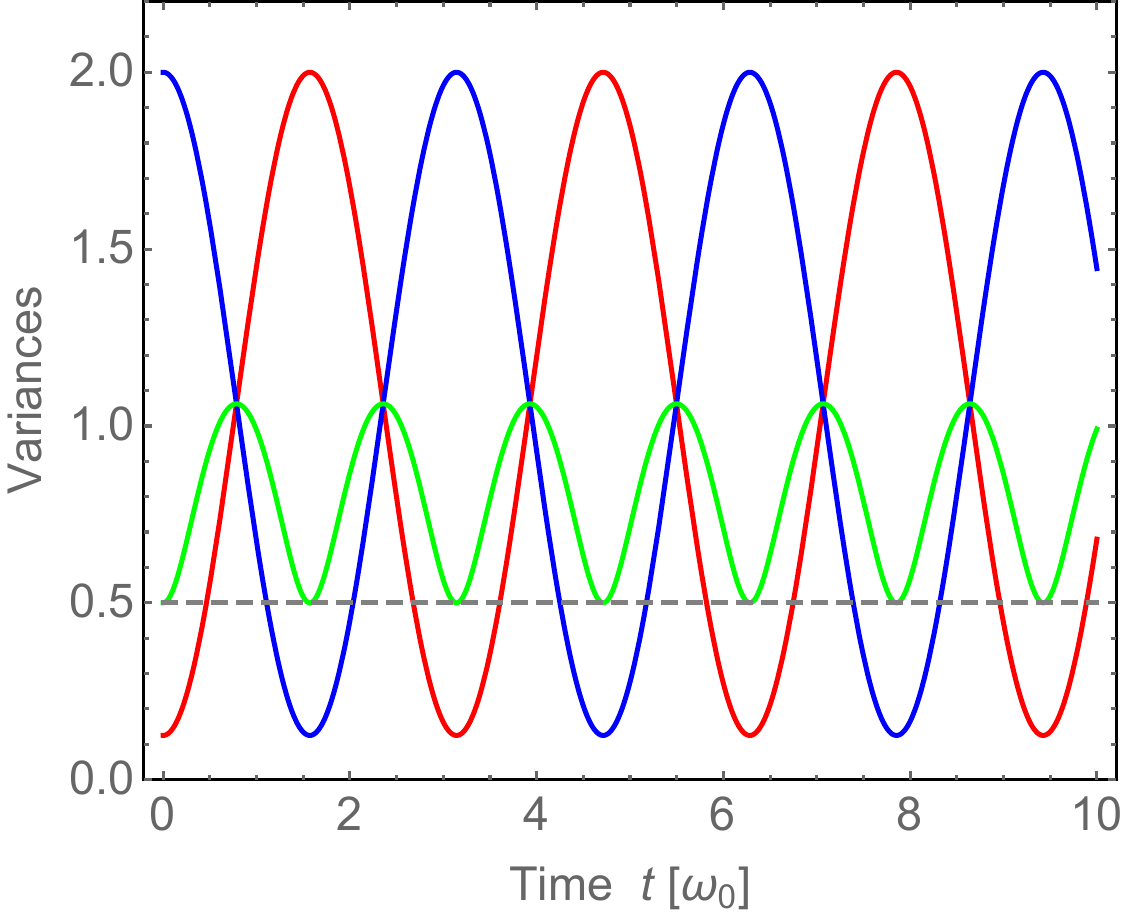}
\caption{Variances vs time for a squeezed state with squeezing parameter $\sq=4$. Red curve is $\sigma_x^2(m\omega_0/\hbar$), blue curve is $\sigma_p^2/(\hbar m\omega_0)$, and green curve is $\sigma_x\,\sigma_p/\hbar$.}
\label{VarianceVsTime}
\end{figure}

The mean of position and momentum is as usual: $\langle x\rangle = x_0(t),\,\langle p\rangle = p_0(t)$. However, the variances are not the usual values from the ground state. They are given by
\ba
&&\sigma_x^2={\hbar\over 2m\omega_0}\!\left(\sq^{-1}\!\cos^2(\omega_0t)+\sq\sin^2(\omega_0t)\right)\,\\
&&\sigma_p^2={\hbar m\omega_0\over 2}\!\left(\sq\cos^2(\omega_0t)+\sq^{-1}\!\sin^2(\omega_0t)\right)\,\,\,\,\,\,\,\,\,\,
\ea
These are plotted in Figure \ref{VarianceVsTime} for the case of $\sq=4$. The product of the standard deviations does not saturate the uncertainty limit. Instead it is given by
\be
\sigma_x\,\sigma_p={\hbar\over2}\sqrt{1+6\sq^2+\sq^4-(\sq^2-1)^2\cos(4\omega_0t)\over8\sq^2}
\ee
This oscillates between the minimum and maximum values of
\be
(\sigma_x\,\sigma_p)_{min}={\hbar\over2},\,\,\,\,\,\,(\sigma_x\,\sigma_p)_{max}={\hbar\over4}\left(\sq^{-1}+\sq\right)
\ee
So by either taking $\sq\gg1$ or $\sq\ll1$ we obtain very large oscillations in the variances.

\subsection{Correlations in Time}

Here we report on the correlation functions in time. To compute this we operate in the Heisenberg picture with
\ba
&&\hat{x}(t) = \hat{x}(0)\,\cos(\omega_0 t)+{\hat{p}(0)\over m\omega_0}\sin(\omega_0t)\\
&&\hat{p}(t) = \hat{p}(0)\,\cos(\omega_0 t)-m\omega_0\,\hat{x}(0)\sin(\omega_0t)
\ea
where $\hat{x}(0),\,\hat{p}(0)$ are standard operators evaluated at $t=0$.
We then define the temporal correlation function as
\be
\xi(t,t') = \langle \hat{x}(t)\,\hat{x}(t')\rangle -x_0(t)x_0(t')
\ee
By computing this expectation value with the above Heisenberg operator results, we obtain
\ba
\xi(t,t')& = &{\hbar\over 2m\omega_0}\Big{(}\sq^{-1}\cos(\omega_0 t)\cos(\omega_0t') \nonumber\\
&&\,\,\,\,\,\,\,\,\,\,\,\,\,\,\,+\sq\,\sin(\omega_0 t)\sin(\omega_0t') \nonumber\\
&&\,\,\,\,\,\,\,\,\,\,\,\,\,\,\,+i\,\sin(\omega_0(t'-t))\Big{)} 
\label{xiONE}\ea
Note that by directly going from $t$ to $t'$, another way of writing this is
\ba
\xi(t,t') &=&{\hbar \over 2m\omega_0 \mathcal{R}[\sqt(t)]} \Big( \! \cos(\omega_0 (t'-t)) \nonumber\\
&&\,\,\,\,\,\,\,\,\,\,\,\,\,\,\,+ i \, \sqt(t) \sin(\omega_0 (t'-t)) \Big)
\label{xiTWO}\ea
We note that in Eq.~(\ref{xiONE}) this is manifestly invariant under the interchange of $t \leftrightarrow t'$ and complex conjugation, though this fact becomes obscured in Eq.~(\ref{xiTWO}). This information is restored when $\sqt(t)$ is written using Eq.~(\ref{S}).

\begin{thebibliography}{}


\bibitem{Weinberg}
S. Weinberg,
``The Quantum Theory of Fields, Vol. 1,''
Cambridge University Press (1995).


\bibitem{Weinberg:1964ew} 
S.~Weinberg,
``Photons and Gravitons in s Matrix Theory: Derivation of Charge Conservation and Equality of Gravitational and Inertial Mass,''
Phys.\ Rev.\  {\bf 135}, B1049 (1964).
%
\bibitem{Weinberg:1965rz}
S.~Weinberg,
``Photons and gravitons in perturbation theory: Derivation of Maxwell's and Einstein's equations,''
Phys. Rev. \textbf{138}, B988-B1002 (1965).
%
\bibitem{Deser:1969wk}
S.~Deser,
``Selfinteraction and gauge invariance,''
Gen. Rel. Grav. \textbf{1}, 9-18 (1970)
[arXiv:gr-qc/0411023 [gr-qc]].
%
\bibitem{Feynman}
R.~P.~Feynman, F.~B.~Morinigo, W.~G.~Wagner, B.~Hatfield, 
``Feynman Lectures on Gravitation," 
Addison-Wesley (1995). 
%
\bibitem{Hertzberg:2016djj}
M.~P.~Hertzberg,
``Constraints on Gravitation from Causality and Quantum Consistency,''
Adv. High Energy Phys. \textbf{2018}, 2657325 (2018)
[arXiv:1610.03065 [hep-th]].
%
\bibitem{Hertzberg:2017abn}
M.~P.~Hertzberg and M.~Sandora,
``General Relativity from Causality,''
JHEP \textbf{09}, 119 (2017)
[arXiv:1702.07720 [hep-th]].
%
\bibitem{Hertzberg:2017nzl}
M.~P.~Hertzberg and M.~Sandora,
``Special Relativity from Soft Gravitons,''
Phys. Rev. D \textbf{96}, no.8, 084048 (2017)
[arXiv:1704.05071 [hep-th]].
%
\bibitem{Hertzberg:2020yzl}
M.~P.~Hertzberg, J.~A.~Litterer and M.~Sandora,
``Symmetries from locality. II. Gravitation and Lorentz boosts,''
Phys. Rev. D \textbf{102}, no.2, 025023 (2020)
[arXiv:2005.01744 [hep-th]].
%
\bibitem{Hertzberg:2020gxu}
M.~P.~Hertzberg and J.~A.~Litterer,
``Symmetries from locality. III. Massless spin-2 gravitons and time translations,''
Phys. Rev. D \textbf{102}, no.8, 085007 (2020)
[arXiv:2008.06510 [hep-th]].



\bibitem{LIGOScientific:2016aoc}
B.~P.~Abbott \textit{et al.} [LIGO Scientific and Virgo],
``Observation of Gravitational Waves from a Binary Black Hole Merger,''
Phys. Rev. Lett. \textbf{116} (2016) no.6, 061102
[arXiv:1602.03837 [gr-qc]].

\bibitem{LIGOScientific:2016lio}
B.~P.~Abbott \textit{et al.} [LIGO Scientific and Virgo],
``Tests of general relativity with GW150914,''
Phys. Rev. Lett. \textbf{116}, no.22, 221101 (2016)
[erratum: Phys. Rev. Lett. \textbf{121}, no.12, 129902 (2018)]
[arXiv:1602.03841 [gr-qc]].



\bibitem{LIGOScientific:2016gtq}
B.~P.~Abbott \textit{et al.} [LIGO Scientific and Virgo],
``Characterization of transient noise in Advanced LIGO relevant to gravitational wave signal GW150914,''
Class. Quant. Grav. \textbf{33}, no.13, 134001 (2016)
[arXiv:1602.03844 [gr-qc]].

\bibitem{LIGO:2021ppb}
D.~Davis \textit{et al.} [LIGO],
``LIGO detector characterization in the second and third observing runs,''
Class. Quant. Grav. \textbf{38}, no.13, 135014 (2021)
[arXiv:2101.11673 [astro-ph.IM]].

\bibitem{McCuller:2021mbn}
L.~McCuller, S.~E.~Dwyer, A.~C.~Green, H.~Yu, L.~Barsotti, C.~D.~Blair, D.~D.~Brown, A.~Effler, M.~Evans and A.~Fernandez-Galiana, \textit{et al.}
``LIGO\textquoteright{}s quantum response to squeezed states,''
Phys. Rev. D \textbf{104}, no.6, 062006 (2021)
[arXiv:2105.12052 [physics.ins-det]].




\bibitem{Rothman:2006fp}
T.~Rothman and S.~Boughn,
``Can gravitons be detected?,''
Found. Phys. \textbf{36}, 1801-1825 (2006)
[arXiv:gr-qc/0601043 [gr-qc]].

\bibitem{Dyson:2013hbl}
F.~Dyson,
``Is a graviton detectable?,''
Int. J. Mod. Phys. A \textbf{28}, 1330041 (2013)



\bibitem{Grishchuk:1993ds}
L.~P.~Grishchuk,
``Quantum effects in cosmology,''
Class. Quant. Grav. \textbf{10}, 2449-2478 (1993)
[arXiv:gr-qc/9302036 [gr-qc]].



\bibitem{Allali:2020ttz}
I.~Allali and M.~P.~Hertzberg,
``Gravitational Decoherence of Dark Matter,''
JCAP \textbf{07}, 056 (2020)
[arXiv:2005.12287 [gr-qc]].

\bibitem{Allali:2020shm}
I.~J.~Allali and M.~P.~Hertzberg,
``Decoherence from General Relativity,''
Phys. Rev. D \textbf{103}, no.10, 104053 (2021)
[arXiv:2012.12903 [gr-qc]].

\bibitem{Allali:2021puy}
I.~J.~Allali and M.~P.~Hertzberg,
``General Relativistic Decoherence with Applications to Dark Matter Detection,''
Phys. Rev. Lett. \textbf{127}, no.3, 031301 (2021)
[arXiv:2103.15892 [gr-qc]].



\bibitem{Albrecht:1992kf}
A.~Albrecht, P.~Ferreira, M.~Joyce and T.~Prokopec,
``Inflation and squeezed quantum states,''
Phys. Rev. D \textbf{50}, 4807-4820 (1994)
[arXiv:astro-ph/9303001 [astro-ph]].

\bibitem{Polarski:1995jg}
D.~Polarski and A.~A.~Starobinsky,
``Semiclassicality and decoherence of cosmological perturbations,''
Class. Quant. Grav. \textbf{13}, 377-392 (1996)
[arXiv:gr-qc/9504030 [gr-qc]].



\bibitem{Parikh:2020nrd}
M.~Parikh, F.~Wilczek and G.~Zahariade,
``The Noise of Gravitons,''
Int. J. Mod. Phys. D \textbf{29}, no.14, 2042001 (2020)
[arXiv:2005.07211 [hep-th]].

\bibitem{Parikh:2020kfh}
M.~Parikh, F.~Wilczek and G.~Zahariade,
``Quantum Mechanics of Gravitational Waves,''
Phys. Rev. Lett. \textbf{127}, no.8, 081602 (2021)
[arXiv:2010.08205 [hep-th]].

\bibitem{Parikh:2020fhy}
M.~Parikh, F.~Wilczek and G.~Zahariade,
``Signatures of the quantization of gravity at gravitational wave detectors,''
Phys. Rev. D \textbf{104}, no.4, 046021 (2021)
[arXiv:2010.08208 [hep-th]].

\bibitem{Kanno:2020usf}
S.~Kanno, J.~Soda and J.~Tokuda,
``Noise and decoherence induced by gravitons,''
Phys. Rev. D \textbf{103}, no.4, 044017 (2021)
[arXiv:2007.09838 [hep-th]].

\bibitem{Kanno:2021gpt}
S.~Kanno, J.~Soda and J.~Tokuda,
``Indirect detection of gravitons through quantum entanglement,''
Phys. Rev. D \textbf{104}, no.8, 083516 (2021)
[arXiv:2103.17053 [gr-qc]].

\bibitem{Guerreiro:2021qgk}
T.~Guerreiro, F.~Coradeschi, A.~M.~Frassino, J.~R.~West and E.~Schioppa, Junior.,
``Quantum Signatures in Nonlinear Gravitational Waves,''
[arXiv:2111.01779 [gr-qc]].

\bibitem{Haba:2020jqs}
Z.~Haba,
``State-dependent graviton noise in the equation of geodesic deviation,''
Eur. Phys. J. C \textbf{81}, no.1, 40 (2021)
[arXiv:2009.12306 [gr-qc]].

\bibitem{Zurek:2020ukz}
K.~M.~Zurek,
``On Vacuum Fluctuations in Quantum Gravity and Interferometer Arm Fluctuations,''
[arXiv:2012.05870 [hep-th]].

\bibitem{Verlinde:2019xfb}
E.~P.~Verlinde and K.~M.~Zurek,
``Observational signatures of quantum gravity in interferometers,''
Phys. Lett. B \textbf{822}, 136663 (2021)
[arXiv:1902.08207 [gr-qc]].

\bibitem{Guerreiro:2019vbq}
T.~Guerreiro,
``Quantum Effects in Gravity Waves,''
Class. Quant. Grav. \textbf{37}, no.15, 155001 (2020)
[arXiv:1911.11593 [quant-ph]].

\bibitem{Kuo:1993if}
C.~I.~Kuo and L.~H.~Ford,
``Semiclassical gravity theory and quantum fluctuations,''
Phys. Rev. D \textbf{47}, 4510-4519 (1993)
[arXiv:gr-qc/9304008 [gr-qc]].

\bibitem{Ford:1994cr}
L.~H.~Ford,
``Gravitons and light cone fluctuations,''
Phys. Rev. D \textbf{51}, 1692-1700 (1995)
[arXiv:gr-qc/9410047 [gr-qc]].

\bibitem{Ford:1996qc}
L.~H.~Ford and N.~F.~Svaiter,
``Gravitons and light cone fluctuations. 2: Correlation functions,''
Phys. Rev. D \textbf{54}, 2640-2646 (1996)
[arXiv:gr-qc/9604052 [gr-qc]].

\bibitem{Coradeschi:2021szx}
F.~Coradeschi, A.~M.~Frassino, T.~Guerreiro, J.~R.~West and E.~Schioppa, Junior.,
``Can We Detect the Quantum Nature of Weak Gravitational Fields?,''
Universe \textbf{7}, no.11, 414 (2021)
[arXiv:2110.02542 [gr-qc]].

\bibitem{Cho:2021gvg}
H.~T.~Cho and B.~L.~Hu,
``Quantum Noise of Gravitons and Stochastic Force on Geodesic Separation,''
[arXiv:2112.08174 [gr-qc]].

\bibitem{Li:2022mvy}
D.~Li, V.~S.~H.~Lee, Y.~Chen and K.~M.~Zurek,
``Interferometer response to geontropic fluctuations,''
Phys. Rev. D \textbf{107}, no.2, 024002 (2023)
[arXiv:2209.07543 [gr-qc]].




\end{thebibliography}
\end{document}